# Quantized Photocurrents in Gapless Topological Matter

**Author:** Byunghoon Kim[1,2,3,4*†], Tenzin Norden[4*], Mohammad Yahyavi[5*], Kaustuv Manna[6,7], Tyler A. Cochran[1], Zi-Jia Cheng[1], Xian P. Yang[1], Yu-Xiao Jiang[1], Xiangyu Luo[1], Payman Kazemikhah[1], Areeq Hasan[8], Vladimir N. Strocov[9], Sergey Shilov[10], Ilya Belopolski[11], Claudia Felser[6], Md Shafayat Hossain[12,13,14], Rohit P. Prasankumar[4,15], Guoqing Chang[5†], M. Zahid Hasan[1,3,8,16†], Prashant Padmanabhan[4†]

**Affiliations:**

[1]Laboratory for Topological Quantum Matter and Advanced Spectroscopy, Department of Physics, Princeton University, Princeton, New Jersey 08544, USA

[2]Department of Chemistry, Princeton University, Princeton, New Jersey 08544, USA

[3]Princeton Institute for Science and Technology of Materials, Princeton University, Princeton, New Jersey 08544, USA

[4]Center for Integrated Nanotechnologies, Los Alamos National Laboratory, Los Alamos, New Mexico 87545, USA

[5]Division of Physics and Applied Physics, School of Physical and Mathematical Sciences, Nanyang Technological University, 21 Nanyang Link 637371, Singapore

[6]Max Planck Institute for Chemical Physics of Solids, 01187 Dresden, Germany

[7]Department of Physics, Indian Institute of Technology Delhi, Hauz Khas, New Delhi 110016, India

[8]Department of Applied Physics, Stanford University, Stanford, California 94305, USA

[9]Swiss Light Source, Paul Scherrer Institute, 5232 Villigen, Switzerland

[10]Bruker Optics, Bruker Scientific, LLC, Billerica, Massachusetts 01821, USA

[11]School of Electrical and Electronic Engineering, Nanyang Technological University, 21 Nanyang Link 637371, Singapore

[12]Department of Materials Science and Engineering, University of California, Los Angeles, California 90095, USA

[13]California NanoSystems Institute, University of California, Los Angeles, California 90095, USA

[14]Center for Quantum Science and Engineering, University of California, Los Angeles, California 90095, USA

[15]Deep Science Fund, Intellectual Ventures, Bellevue, Washington 98005, USA
[16]Lawrence Berkeley National Laboratory, Berkeley, California 94720, USA

†Corresponding authors: Byunghoon Kim (bhkim@princeton.edu); Guoqing Chang (guoqing.chang@ntu.edu.sg); M. Zahid Hasan (mzhasan@princeton.edu); Prashant Padmanabhan (prashpad@lanl.gov)

*These authors contributed equally to this work.

**Abstract**

**The quantum Hall effect establishes that topology can fix a material response to integer multiples of fundamental constants when an energy gap isolates the relevant symmetry-protected electronic states. Whether such universal quantization can also emerge in gapless matter, where topological bands coexist with a continuum of metallic excitations, has remained a fundamental question in the field of quantum materials. Chiral topological semimetals provide a unique setting in which to explore this principle; when optical transitions are confined to a single chiral node, the resulting circular photogalvanic effect is predicted to be quantized by the topological charge of the node. In real materials, however, this nonlinear optical phenomenon has remained experimentally elusive, obscured by trivial band transitions, insufficient energy separation between node pairs, and their relative positions with respect to the Fermi level. Here we observe a quantized circular photogalvanic effect in the chiral topological semimetal $Rh_{0.95}Ni_{0.05}Si$. Band engineering via Ni substitution opens a photon-energy window dominated by interband optical transitions at the Γ-point multifold node. This allows circularly polarized near- to mid-infrared pulses to drive a helicity-odd terahertz response that manifests three hallmarks of quantization: a sharp onset, a photon-energy-independent plateau governed by the magnitude of the monopole charge, and an abrupt long-wavelength cutoff imposed by Pauli blocking. Our work thus establishes an all-optical analogue of the quantum Hall effect and a new paradigm to realize topological quantization in gapless matter.**

## Main

Topological phases can exhibit quantized physical response coefficients fixed by global invariants of the electronic wavefunction[1–3]. In gapped systems, this principle is embodied by the quantum Hall effect, in which the topology of the occupied electronic states fixes the Hall conductance in units of $e^2/h$, with an integer multiple equivalent to the Chern number[1–9]. These landmark demonstrations, however, rely on a bulk energy gap that intrinsically protects the response from low-energy excitations. By contrast, establishing an analogous quantized response in bulk gapless systems is hindered by the presence of low-energy quasiparticles and interband couplings, which can obscure or renormalize the measured response[10,11]. Demonstrating quantized responses from gapless systems, therefore, remains a central open challenge in the field of quantum materials[12,13].

Chiral topological semimetals provide a promising platform in which to address this challenge[13–16]. Their electronic structures host symmetry-protected, multifold band crossings whose Berry flux through a surface enclosing the node is quantized, defining an integer monopole charge, $C$ (Fig. 1a)[14,17,18] The circular photogalvanic effect (CPGE), a second-order nonlinear optical response in which circularly polarized light drives a helicity-odd injection current, has been proposed as a direct optical probe of $C$ in gapless systems[18–22]. When the relevant optical transitions are restricted to only a single chiral node, the trace of the CPGE tensor is predicted to take the universal form $\mathrm{Tr}[\beta] = C\pi \frac{e^3}{h^2}$[18,22,23]. This quantization can, in principle, be directly measured through the helicity difference of the terahertz (THz) emission arising from the injection current.

The main difficulty lies in realizing the necessary single-node limit. The relevant photon-energy window must retain interband transitions involving only one chiral node while excluding contributions from the opposite-charge counterpart[18,24]. In prototypical Weyl semimetals, such as TaAs, this separation is precluded by the degenerate energy of the Weyl node pairs[16,25–27]. Chiral multifold semimetals avoid this constraint, as their lack of mirror symmetries allows opposite-chirality nodes to substantially separate in energy, so that one node can in principle be Pauli-blocked entirely[21,22,28]. In this context, transition-metal monosilicide crystals are a promising candidate material family[24,28–32]. Yet stoichiometric compounds, such as RhSi, still do not exhibit the predicted single-node response[30,31]. The Brillouin-zone-center node, where multiple bands converge in a higher-fold degeneracy, lies slightly above the Fermi level. Consequently, the "across-node" transitions from the flat-like band (red plane, Fig. 1b) at the Γ-point are Pauli-blocked (dashed arrow, Fig. 1b) and the response is instead dominated by competing transitions (solid arrows, middle panel of Fig. 1b)[28,33,34]. Beyond this, theoretical and experimental studies have shown that its opposite-charge counterpart at the R-point contributes transitions of opposite sign, obscuring the quantized signature, while the spin-orbit-split bands superimpose a weaker non-quantized background[31,35]. Therefore, the net optical response is a superposition of competing channels rather than the hallmark of a single Berry monopole. As such, despite extensive experimental efforts[30,31,36], unambiguous observation of the predicted quantized response has remained elusive.

Here we isolate and measure the quantized CPGE associated with a single Berry monopole in $Rh_{0.95}Ni_{0.05}Si$. Electron-doping from the Ni substitution of Rh (Fig. 1c) shifts the RhSi-derived band manifold to higher binding energy and places the zone-center multifold node slightly below $E_F$ (right panel of Fig. 1b), as confirmed by density functional theory (DFT) and angle-resolved photoemission spectroscopy (ARPES; Figs. 1d and 1e, respectively). This band engineering leads to a finite photon-energy window in which interband excitation is dominated by the zone-center node[33,37] (right panel of Fig. 1b). When driven by ultrafast near- to mid-infrared pulses, the resulting single-node optical window gives rise to a distinctive THz emission response in which the helicity-odd signal turns on sharply, forms a broad photon-energy-independent plateau with a calibrated magnitude set by the monopole charge, and vanishes at a low-photon-energy Pauli-blocking edge. Crucially, a single node underlies two otherwise independent observables. Its monopole charge dictates the plateau magnitude, while the Pauli blocking of the same transitions that build the plateau sets its cutoff. This link identifies the emission as a Berry-monopole-selected injection current, distinct from a generic nonlinear optical response. First-principles calculations reproduce the full spectral structure and trace it to optical transitions localized near the isolated Γ-point node. Our results show that chemical tuning can enable quantized nonlinear optical responses, providing an all-optical framework for the measurement of topological invariants in bulk gapless matter.

## Results

### Helicity-dependent THz-emission

Ultrafast photocurrent responses in chiral semimetals arise from several linear and nonlinear optical mechanisms, distinguished by their dependence on the pump helicity. Helicity-even mechanisms such as optical rectification, photothermal emission, photo-Dember currents, and certain bulk photovoltaic effects generate THz fields that are symmetric under the exchange of right- (RCP) and left-circularly polarized (LCP) light[19,38–41]. Meanwhile, helicity-odd responses, antisymmetric in the pump helicity, are far less common, particularly in non-magnetic systems. Indeed, in the case of chiral semimetals, they are a defining feature of the CPGE[18,22,29]. To explore these effects, we used THz emission spectroscopy in which near- and mid-infrared pulses were incident at 45° on a [001]-oriented single-crystal of $Rh_{0.95}Ni_{0.05}Si$ (Fig. 2a and Extended Fig. 1)[30,36,42]. This configuration ensures that any longitudinal photocurrent generated along the pump wavevector acquires a finite surface-parallel component, which radiates into the far field (see Supplementary Note 1). Under 2200 nm excitation, RCP and LCP pump pulses yield almost purely $p$-polarized THz waveforms whose profiles are nearly equivalent in magnitude and absolute structure but opposite in polarity (Fig. 2b).

To establish the polarization symmetry, we measured the quarter-wave-plate (QWP) angle dependence of the peak THz field under 2200 nm photoexcitation (Extended Figs. 2a and 2b; see Supplementary Figure 1 for the full wavelength range). This can be modeled with the phenomenological form[43]

$$E(\phi) = A_1 \sin(2\phi) + A_2 \sin(4\phi) + A_3 \cos(4\phi) + A_0, \quad (1)$$

where $A_1 \sin(2\phi)$ represents the CPGE component, $A_2 \sin(4\phi)$ and $A_3 \cos(4\phi)$ represent linear polarization contributions, $A_0$ is the polarization independent term, and $\phi$ is the angle between the fast-axis of the QWP and the horizontal axis of the incident light. By fitting the measured polarization dependence to Eq. (1), we can quantitatively compare the relative amplitudes of the various terms. The $p$-polarized response is dominated by the $A_1$ term, consistent with a helicity-dependent photocurrent that originates from the CPGE (Extended Figs. 2a and 2b). By contrast, the $s$-polarized THz emission remains below the experimental noise level, pointing to negligible contributions from helicity-even responses[44–46].

In the ideal band structure limit and for a fixed pump helicity, the CPGE photocurrent generated from two oppositely charged chiral nodes are expected to be equal and opposite (*e.g.*, $J_{RCP}^{+C} = -J_{RCP}^{-C}$), provided the two nodes contribute with equivalent weight. The same is true for the opposing pump helicity. Therefore, the net CPGE photocurrent would vanish identically under both RCP and LCP pumping. This balance is upset if optical transitions stem from only a single node. To verify that this is the case for $Rh_{0.95}Ni_{0.05}Si$, we measured the THz response while tuning the pump wavelength. We observe that the $p$-polarized emission is confined to a sharply bounded spectral interval (Figs. 2c and 2d). From the high-energy edge, the response rises steeply between 1600 nm and 2000 nm, marking the entry into the interband transition window associated with the $\Gamma$-point multifold node, since transitions associated with the $R$-point node are strongly Pauli-blocked (Fig. 1d). Between approximately 2000 and 5300 nm, the peak THz field varies minimally. This wavelength-independent response arises from the Ni substitution that shifts the $\Gamma$-point multifold node below $E_F$, with a binding energy of approximately 200 meV (Figs. 1d and 1e). The accessible interband transitions over this wavelength range remain confined to the Γ-point band manifold, while the opposite-charge node near the R-point remains excluded by Pauli blocking due to the band-energy mismatch. At wavelengths longer than approximately 6000 nm, the emission under circularly polarized photoexcitation abruptly quenches, consistent with Pauli blocking of the Γ-point node beyond this wavelength. The sharp onset, minimal wavelength dependence across the intermediate regime, and abrupt long-wavelength cutoff suggest an interband transition region highly localized in $k$-space around the Γ-point multifold node, a consequence of the doping-engineered band alignment.

**Quantized CPGE response from a single-monopole**

Having established the spectrally selective optical window, we explore whether the helicity-dependent THz emission manifests a quantized response resulting from the isolated Γ-point monopole charge. This is accomplished by benchmarking the emitted field against the THz emission from optical rectification in a (110) ZnTe reference measured in the same experimental geometry (see Supplementary Note 2). This calibration yields the response product $\mathrm{Tr}[\beta]\tau$, where $\tau$ denotes the relaxation time of the helicity-dependent CPGE photocurrent, which converts the injection rate into a transient current amplitude, distinct from a hot-carrier energy-relaxation lifetime[18,30,36]. The emitted field is proportional to this product, so the measurement alone does not separate its two factors. We therefore fix $\tau$ by requiring the calculated CPGE magnitude to reproduce the measured one (Supplementary Note 3), which places the response on the quantized scale $\mathrm{Tr}[\beta]/\beta_0$, with $\beta_0 = \pi e^3/h^2$, on which an isolated Berry monopole charge $C$ satisfies $\mathrm{Tr}[\beta] = C\beta_0$. The spectral boundaries are independent of this step. Thus, the calibrated CPGE trace can be compared directly with the integer charge $C$ expected for a single Berry monopole[18,24].

As shown in Fig. 3, $Rh_{0.95}Ni_{0.05}Si$ exhibits a broad mid-infrared plateau between approximately 2000 nm to 5300 nm, over which the calibrated response ($\mathrm{Tr}[\beta]/\beta_0$) results in an average CPGE trace, $C_{\mathrm{exp}}$ ≃2.39, slightly above the ideal charge $C = 2$ of the Γ-point multifold node. We note that spin-orbit coupling splits the Γ-point node into a fourfold chiral fermion, which carries Chern number +4 at the Fermi level[22,47]. This splitting is confined to a few tens of meV, which is below the resolution limit of our previous ARPES studies on the same crystals[28,33,37] and an order of magnitude smaller than the photon energies of plateau. Nevertheless, these residual spin-orbit effects enter indirectly, through the orbital and spin texture of the multifold wavefunctions, leading to the observed enhancement of $C_{\mathrm{exp}}$. As discussed in the following section, this feature is reproduced by our first-principles calculation once the realistic band structure of the engineered node is included.

Figure 3 allows us to identify two unambiguous signatures of quantized optical responses in $Rh_{0.95}Ni_{0.05}Si$: a calibrated response that approaches the nearly quantized CPGE value and a finite onset-plateau-cutoff window in photon energy. Conventional photocurrent contributions can be reshaped in amplitude by absorption, penetration depth, and optical out-coupling. However, no single mechanism simultaneously pins the response to the near-quantized magnitude set by the topological charge, imposes sharp spectral boundaries, and enforces helicity-odd symmetry.

**Band-resolved origin of the quantized CPGE**

To confirm the band-resolved origin of the $\beta\tau$ plateau, we computed the CPGE response of $Rh_{0.95}Ni_{0.05}Si$ from first principles[18,33,48]. The calculation is parameter free, relying only on the band structure obtained from the DFT manifold, which was tuned to match our ARPES measurements, and with the Fermi level fixed at the position established by the experimental results (see Figs. 4a and 4c and Methods). In the isolated interband-transition regime, where the optical linewidth is small compared with the photon energy and the relevant band separations, the CPGE generates an injection current at a rate governed by the second-order response tensor $\beta_{ij}(\omega)$,

$$\frac{dj_i}{dt} = \beta_{ij}(\omega)\,[\mathbf{E}(\omega) \times \mathbf{E}^*(\omega)]_j, \tag{2}$$

where $\mathbf{E}(\omega) = \mathbf{E}^*(-\omega)$ is the electric field of the incident light, and $\beta_{ij}(\omega)$ is the second-order CPGE tensor[18]. Within the independent-particle approximation, $\beta_{ij}(\omega)$ admits the length-gauge form

$$\beta_{ij}(\omega) = \frac{-i\pi e^3}{\hbar^2}\,\epsilon_{jab} \int_{\mathrm{BZ}} \frac{d^3k}{(2\pi)^3} \sum_{n,m} f_{nm}\, \Delta^i_{mn}\, r^a_{nm}\, r^b_{mn}\, \delta(\omega_{mn} - \omega), \tag{3}$$

where the Fermi-Dirac distribution function $f_{nm} = f_n - f_m$ enforces Pauli blocking between bands $n$ and $m$, $\Delta^i_{mn}$ is the interband velocity difference, and $r^a_{nm}$ is the interband Berry connection. Here $\epsilon_{jab}$ is summed over the repeated indices, and we adopted the convention in which $\beta_{ij}(\omega)$ is real; it differs by a factor of $i$ from the definition[18], where the CPGE tensor is purely imaginary. The observable of interest is the Tr $[\beta(\omega)] \equiv \sum_i \beta_{ii}(\omega)$, which approaches the single-node quantized value $C\beta_0$, whenever the optical window is dominated by a single isolated band crossing of monopole charge $C$. In the cubic chiral point group of $Rh_{0.95}Ni_{0.05}Si$, the three diagonal components contribute equally to $\beta(\omega)$ by symmetry (see Methods for their explicit forms).

The first-principles calculation of $\beta(\omega)$ is overlaid on the experimental $\beta\tau$ response as the dashed line in Fig. 3. Without any phenomenological tuning parameters, the theory shows excellent agreement with the spectral position of the observed plateau, its abrupt onset, and its sudden quenching at the long-wavelength Pauli edge. Its plateau value, $C_{\mathrm{DFT}} \simeq 2.17$, closely tracks the measured $C_{\mathrm{exp}}$ and both lie modestly above the ideal topological invariant $C = 2$ expected for the chemically isolated Γ-point multifold transition.

To trace this agreement to specific interband channels, we decompose $\beta(\omega)$ into contributions resolved in momentum space and in band index (Figs. 4b and 4d). Across the entire plateau window, the momentum-resolved integrand of $\beta(\omega)$ is concentrated in a small region of the Brillouin zone surrounding Γ. As such, the plateau is not built from a balance of transitions across $k$-space but arises from a region localized around a single multifold node. The Ni-induced band-shift places the smallest optically allowed transition of this node at the photon energy where the THz emission signal quenches as shown in Figs. 2 and 3, identifying the long-wavelength limit due to the Pauli-blocking threshold associated with the engineered band alignment. As $\omega$ increases beyond the plateau, transitions near the R-point multifold node enter the optical window. The R-point node is the opposite-chirality partner of the Γ-point multifold node, and chemical engineering of the Fermi level is designed precisely to keep its transitions out of the plateau window; its oppositely signed CPGE contribution suppresses and ultimately reverses $\beta(\omega)$, consistent with the vanishing net Berry-monopole charge required across the Brillouin zone. This picture is corroborated by the calculated evolution of $\beta(\omega)$ under a rigid shift of the band manifold (Extended Data Fig. 3). At the intrinsic alignment, the response reverses sign within the mid-infrared regime, whereas a downward shift matching the Ni-induced doping isolates the positive single-node plateau and its long-wavelength cutoff.

**Conclusion**

In summary, we have realized the long-sought quantized CPGE response in a gapless topological semimetal. Our experiments and first-principles calculations on $Rh_{0.95}Ni_{0.05}Si$ reveal the hallmark onset-plateau-cutoff in the trace of the CPGE tensor, with a calibrated magnitude that approaches the monopole charge. Here the response is engineered by tuning the chemical potential until a single multifold node governs the optical window, while still inheriting the band topology of the parent compound RhSi. The same principle, if not the same chemistry, extends across the broader set of chiral multifold semimetals. Across this family, the higher-fold crossings take varied forms. AlPt and related compounds host four- and sixfold fermions that act as Berry monopoles of larger Chern number[21,24,49], while PdGa and PtGa offer maximal-charge platforms in which crystal handedness determines the sign of the topological charge[50,51]. An all-optical probe is well suited to resolve this diversity. The helicity-odd CPGE photocurrent and the associated THz response directly reads the multifold charge from the bulk, with a signal that grows with the degeneracy of the crossing and the strength of spin-orbit coupling. At fixed chiral symmetry, tuning the chemical potential offers a systematic route to bring each node into its own single-monopole window. This establishes a bulk-sensitive route to quantifying multifold charge that complements the counting of Fermi arcs in photoemission. Moreover, because the CPGE originates from bulk interband transitions, it determines the charge without invoking bulk-boundary correspondence, and is insensitive to the crystals surface termination and preparation. Ultimately, our work extends ultrafast nonlinear photocurrents from a qualitative signature of crystal symmetry into a quantitative, contact-free measurement of integer topological charge. More broadly, the combination of chemical engineering and broadband THz emission spectroscopy offers a general paradigm for realizing and identifying topologically fixed optical responses in Weyl, multifold, and other gapless systems.

## Reference

1. Thouless, D. J., Kohmoto, M., Nightingale, M. P. & den Nijs, M. Quantized Hall conductance in a two-dimensional periodic potential. *Physical review letters* **49**, 405 (1982).
2. Kane, C. L. & Mele, E. J. Z 2 topological order and the quantum spin Hall effect. *Physical review letters* **95**, 146802 (2005).
3. Hasan, M. Z. & Kane, C. L. Colloquium: topological insulators. *Reviews of modern physics* **82**, 3045–3067 (2010).
4. Bernevig, B. A., Hughes, T. L. & Zhang, S.-C. Quantum spin Hall effect and topological phase transition in HgTe quantum wells. *science* **314**, 1757–1761 (2006).
5. Haldane, F. D. M. Model for a quantum Hall effect without Landau levels: Condensed-matter realization of the" parity anomaly". *Physical review letters* **61**, 2015 (1988).
6. Laughlin, R. B. Anomalous quantum Hall effect: An incompressible quantum fluid with fractionally charged excitations. *Physical Review Letters* **50**, 1395 (1983).
7. Kane, C. L. & Mele, E. J. Quantum spin Hall effect in graphene. *Physical review letters* **95**, 226801 (2005).
8. Tsui, D. C., Stormer, H. L. & Gossard, A. C. Two-dimensional magnetotransport in the extreme quantum limit. *Physical Review Letters* **48**, 1559 (1982).
9. Konig, M. *et al.* Quantum spin Hall insulator state in HgTe quantum wells. *Science* **318**, 766–770 (2007).
10. Avdoshkin, A., Kozii, V. & Moore, J. E. Interactions remove the quantization of the chiral photocurrent at Weyl points. *Physical review letters* **124**, 196603 (2020).
11. Vazifeh, M. M. & Franz, M. Electromagnetic response of Weyl semimetals. *Physical review letters* **111**, 027201 (2013).
12. Wan, X., Turner, A. M., Vishwanath, A. & Savrasov, S. Y. Topological semimetal and Fermi-arc surface states in the electronic structure of pyrochlore iridates. *Physical Review B—Condensed Matter and Materials Physics* **83**, 205101 (2011).
13. Armitage, N. P., Mele, E. J. & Vishwanath, A. Weyl and Dirac semimetals in three-dimensional solids. *Reviews of Modern Physics* **90**, 015001 (2018).
14. Burkov, A. A. & Balents, L. Weyl semimetal in a topological insulator multilayer. *Physical review letters* **107**, 127205 (2011).
15. Hasan, M. Z. *et al.* Weyl, Dirac and high-fold chiral fermions in topological quantum matter.

*Nature Reviews Materials* **6**, 784–803 (2021).

16. Xu, S.-Y. *et al.* Discovery of a Weyl fermion semimetal and topological Fermi arcs. *Science* **349**, 613–617 (2015).
17. Hasan, M. Z., Xu, S.-Y., Belopolski, I. & Huang, S.-M. Discovery of Weyl fermion semimetals and topological Fermi arc states. *Annual Review of Condensed Matter Physics* **8**, 289–309 (2017).
18. De Juan, F., Grushin, A. G., Morimoto, T. & Moore, J. E. Quantized circular photogalvanic effect in Weyl semimetals. *Nature communications* **8**, 15995 (2017).
19. Moore, J. E. & Orenstein, J. Confinement-induced Berry phase and helicity-dependent photocurrents. *Physical review letters* **105**, 026805 (2010).
20. Sodemann, I. & Fu, L. Quantum nonlinear Hall effect induced by Berry curvature dipole in time-reversal invariant materials. *Physical review letters* **115**, 216806 (2015).
21. Chang, G. *et al.* Topological quantum properties of chiral crystals. *Nature materials* **17**, 978–985 (2018).
22. Flicker, F. *et al.* Chiral optical response of multifold fermions. *Physical Review B* **98**, 155145 (2018).
23. Orenstein, J. *et al.* Topology and symmetry of quantum materials via nonlinear optical responses. *Annual Review of Condensed Matter Physics* **12**, 247–272 (2021).
24. Bradlyn, B. *et al.* Beyond Dirac and Weyl fermions: Unconventional quasiparticles in conventional crystals. *Science* **353**, aaf5037 (2016).
25. Weng, H., Fang, C., Fang, Z., Bernevig, B. A. & Dai, X. Weyl semimetal phase in noncentrosymmetric transition-metal monophosphides. *Physical Review X* **5**, 011029 (2015).
26. Ma, Q. *et al.* Direct optical detection of Weyl fermion chirality in a topological semimetal. *Nature Physics* **13**, 842–847 (2017).
27. Osterhoudt, G. B. *et al.* Colossal mid-infrared bulk photovoltaic effect in a type-I Weyl semimetal. *Nature materials* **18**, 471–475 (2019).
28. Sanchez, D. S. *et al.* Topological chiral crystals with helicoid-arc quantum states. *Nature* **567**, 500–505 (2019).
29. Chang, G. *et al.* Unconventional photocurrents from surface Fermi arcs in topological chiral semimetals. *Physical review letters* **124**, 166404 (2020).
30. Rees, D. *et al.* Helicity-dependent photocurrents in the chiral Weyl semimetal RhSi. *Sci. Adv.*

**6**, eaba0509 (2020).

31. Ni, Z. *et al.* Linear and nonlinear optical responses in the chiral multifold semimetal RhSi. *npj Quantum Mater.* **5**, 96 (2020).
32. Le, C. & Sun, Y. Topology and symmetry of circular photogalvanic effect in the chiral multifold semimetals: a review. *J. Phys.: Condens. Matter* **33**, 503003 (2021).
33. Cochran, T. A. *et al.* Visualizing Higher-Fold Topology in Chiral Crystals. *Phys. Rev. Lett.* **130**, 066402 (2023).
34. Wang, J. *et al.* Quantifying quasiparticle chirality in a chiral topological semimetal.
35. de Juan, F. *et al.* Difference frequency generation in topological semimetals. *Physical Review Research* **2**, 012017 (2020).
36. Ni, Z. *et al.* Giant topological longitudinal circular photo-galvanic effect in the chiral multifold semimetal CoSi. *Nature communications* **12**, 154 (2021).
37. Sanchez, D. S. *et al.* Tunable topologically driven Fermi arc van Hove singularities. *Nature Physics* **19**, 682–688 (2023).
38. Sturman, B. & Fridkin, V. *Photovoltaic and Photo-Refractive Effects in Noncentrosymmetric Materials*. (Routledge, 2021).
39. Sirica, N. *et al.* Tracking ultrafast photocurrents in the Weyl semimetal TaAs using THz emission spectroscopy. *Physical review letters* **122**, 197401 (2019).
40. Ganichev, S. D. & Prettl, W. Spin photocurrents in quantum wells. *Journal of physics: Condensed matter* **15**, R935–R983 (2003).
41. Pettine, J. *et al.* Ultrafast terahertz emission from emerging symmetry-broken materials. *Light: Science & Applications* **12**, 133 (2023).
42. Papaioannou, E. Th. & Beigang, R. THz spintronic emitters: a review on achievements and future challenges. *Nanophotonics* **10**, 1243–1257 (2021).
43. McIver, J. W., Hsieh, D., Steinberg, H., Jarillo-Herrero, P. & Gedik, N. Control over topological insulator photocurrents with light polarization. *Nature nanotechnology* **7**, 96–100 (2012).
44. Nastos, F. & Sipe, J. E. Optical rectification and shift currents in GaAs and GaP response: Below and above the band gap. *Physical Review B—Condensed Matter and Materials Physics* **74**, 035201 (2006).
45. Boland, J. L. *et al.* Narrowband, angle-tunable, helicity-dependent terahertz emission from

nanowires of the topological Dirac semimetal $Cd_3As_2$. *ACS photonics* **10**, 1473–1484 (2023).

46. Chen, Z. *et al.* Defect-induced helicity dependent terahertz emission in Dirac semimetal $PtTe_2$ thin films. *Nature Communications* **15**, 2605 (2024).
47. Chang, G. *et al.* Unconventional chiral fermions and large topological Fermi arcs in RhSi. *Physical review letters* **119**, 206401 (2017).
48. Sipe, J. E. & Shkrebtii, A. I. Second-order optical response in semiconductors. *Physical Review B* **61**, 5337 (2000).
49. Schröter, N. B. *et al.* Chiral topological semimetal with multifold band crossings and long Fermi arcs. *Nature Physics* **15**, 759–765 (2019).
50. Sessi, P. *et al.* Handedness-dependent quasiparticle interference in the two enantiomers of the topological chiral semimetal PdGa. *Nature communications* **11**, 3507 (2020).
51. Schröter, N. B. *et al.* Observation and control of maximal Chern numbers in a chiral topological semimetal. *Science* **369**, 179–183 (2020).
52. Strocov, V. N. *et al.* Soft-X-ray ARPES facility at the ADRESS beamline of the SLS: concepts, technical realisation and scientific applications. *Synchrotron Radiation* **21**, 32–44 (2014).

**Methods**

**Single crystal synthesis.** Single crystals of $Rh_{0.95}Ni_{0.05}Si$ were synthesized using a vertical Bridgman melt-growth technique. High-purity Rh, Ni, and Si were combined according to the nominal composition $Rh_{0.95}Ni_{0.05}Si$, with a slight excess of Si added to facilitate flux-assisted crystal growth. The starting materials were first arc-melted under an argon atmosphere to obtain a homogeneous polycrystalline precursor. The precursor was then crushed into powder, loaded into a sharp-bottom alumina crucible, and sealed inside a tantalum ampule under argon.

For the Bridgman growth, the sealed ampule was heated to 1550 °C at a rate of 200 °C $h^{-1}$ and held at this temperature for 10 h. It was then slowly translated through the furnace temperature gradient toward the cold zone until the temperature reached 1100 °C, using a pulling rate of 0.8 mm $h^{-1}$. The growth temperature was monitored with a thermocouple positioned near the bottom of the ampule. The resulting single crystals had typical dimensions of approximately 9 mm in length and 6 mm in diameter.

**Angle-resolved photoemission spectroscopy.** Prior to photoemission measurements, (001) polished single crystals underwent cyclic in situ preparation in ultra-high vacuum (UHV), consisting of Ar ion sputtering at 500-1000 eV for 5-30 minutes followed by annealing at 550-585°C (based on an assumed sample emissivity of 0.45) for 30-45 minutes. Soft X-ray angle-resolved photoemission spectroscopy (SXARPES) measurements on $Rh_{0.95}Ni_{0.05}Si$ were performed at beamline BL25SU of SPring-8, using incident photon energies between 400-750 eV and a Scienta DA30 deflection-based hemispherical analyzer. All ARPES measurements on this sample were conducted at 21-25 K under UHV conditions. The use of soft X-ray energies extends the photoelectron mean-free path, which, via the Heisenberg uncertainty principle, enhances the intrinsic momentum resolution of the experiment-a particularly important advantage for three-dimensional materials. For RhSi, SXARPES measurements were carried out at the ADRESS beamline of the Swiss Light Source[52], again with incident photon energies in the 400-750 eV range, using a Specs PHOIBOS-150 analyzer. These measurements were performed below 20 K under UHV conditions.

**THz emission spectroscopy.** THz emission spectroscopy was performed at room temperature (298 K) in a Nitrogen environment with relative humidity below 1%. Tunable near- and mid-infrared pump pulses were generated from an ultrafast Ti:sapphire amplifier using an optical parametric amplifier and difference-frequency generation, covering wavelengths from 1.6 to 15 μm. The pump beam was incident on the [001]-oriented $Rh_{0.95}Ni_{0.05}Si$ crystal at 45° relative to the surface normal. The pump polarization was controlled using QWPs, allowing right- and left-circularly polarized excitation as well as full QWP-angle scans. The emitted THz radiation was collected and refocused using off-axis unprotected gold parabolic mirrors, passed through a wire-grid polarizer to select the electric-field component in the plane of incidence, and detected by electro-optic sampling in a ZnTe (110) crystal with a synchronized 800 nm gate pulse. Absolute calibration of the CPGE response was performed using the THz emission from a ZnTe (110) crystal generated via optical-rectification, measured in the same optical geometry. Details of the calibration procedure and uncertainty analysis are given in Supplementary Note 2.

**Optical conductivity and Kramers-Kronig analysis.** Normal-incidence reflectivity, $R(\omega)$, was measured using Bruker VERTEX 70v (650 to 8000 $cm^{-1}$) and VERTEX 80v (30 to 600 $cm^{-1}$) Fourier-transform infrared (FTIR) spectrometers. Absolute reflectivity was obtained by referencing a gold film, eliminating uncertainties associated with sample positioning and surface reflectance. The measured reflectivity spectrum was extended toward $\omega \rightarrow 0$ using a metallic Hagen-Rubens form extrapolation, while the high-frequency region was extrapolated using the standard free-electron asymptotic behavior. The extended reflectivity was then used as input for the Kramers-Kronig transformation to obtain the complex optical constants, including $\sigma_1(\omega)$, $\sigma_2(\omega)$, $n(\omega)$, and $k(\omega)$. These were used to determine the pump penetration depth and optical boundary factors required for the absolute CPGE calibration. The same optical-conductivity data were also used to estimate the equilibrium Drude scattering time $\tau_D$ by fitting the low-frequency $\sigma_1(\omega)$ response with a Drude model

**CPGE tensor: definition and component-resolved form.** The CPGE is the second-order dc nonlinear optical response induced by circularly polarized light in nonmagnetic chiral materials. In the clean limit, where the interband relaxation rate is small compared with both the band gap and the photon frequency, the CPGE current is described by

$$\frac{dj_i}{dt} = \beta_{ij}(\omega)[\mathbf{E}(\omega) \times \mathbf{E}^*(\omega)]_j, \tag{M1}$$

where $\mathbf{E}(\omega) = \mathbf{E}^*(-\omega)$ is the electric field of the incident light and $\beta_{ij}(\omega)$ is the second-order CPGE tensor.

Within the independent-particle approximation, the CPGE tensor takes the form

$$\beta_{ij}(\omega) = \frac{-i\pi e^3}{\hbar^2} \epsilon_{jab} \int_{\mathrm{BZ}} \frac{d^3k}{(2\pi)^3} \sum_{n,m} f_{nm}\, \Delta^i_{mn}\, r^a_{nm}\, r^b_{mn}\, \delta(\omega_{mn} - \omega), \tag{M2}$$

where the integration runs over the Brillouin zone, $\hbar\omega_{mn} = E_m - E_n$ is the energy difference between bands $m$ and $n$, $f_n$ is the Fermi-Dirac distribution function, and $f_{nm} = f_n - f_m$. The quantity $\Delta^i_{mn} = (1/\hbar)\ \partial_{k_i} E_{mn}$ is the interband velocity difference, and $r^\beta_{nm} = i\,\langle u_{n\mathbf{k}} \mid \partial_{k_\beta} \mid u_{m\mathbf{k}}\rangle$ is the interband Berry connection, corresponding to the optical transition matrix element between Bloch states. The Levi-Civita symbol $\epsilon_{jab}$ encodes the polarization selection rule of the circularly polarized drive.

The quantity of central interest in the main text is the trace of $\beta_{ij}(\omega)$,

$$\beta(\omega) = \mathrm{Tr}\left[\beta_{ij}(\omega)\right] = \beta_{xx}(\omega) + \beta_{yy}(\omega) + \beta_{zz}(\omega). \tag{M3}$$

The three diagonal components entering the trace are obtained from Eq. (M2) by setting $i = j$. The antisymmetric sum over $(a, b)$ yields $2i\ \mathrm{Im}[r^a_{nm}\, r^b_{mn}]$,

$$\beta_{xx}(\omega) = \frac{2\pi e^3}{\hbar^2} \int \frac{d^3k}{(2\pi)^3} \sum_{n,m} f_{nm}\, \Delta^x_{mn}\, \mathrm{Im}\left[r^y_{nm}\, r^z_{mn}\right] \delta(\omega_{mn} - \omega),$$

$$\beta_{yy}(\omega) = \frac{2\pi e^3}{\hbar^2} \int \frac{d^3k}{(2\pi)^3} \sum_{n,m} f_{nm}\, \Delta^y_{mn}\, \mathrm{Im}\left[r^z_{nm}\, r^x_{mn}\right] \delta(\omega_{mn} - \omega),$$

$$\beta_{zz}(\omega) = \frac{2\pi e^3}{\hbar^2} \int \frac{d^3k}{(2\pi)^3} \sum_{n,m} f_{nm}\, \Delta^z_{mn}\, \mathrm{Im}\left[r^x_{nm}\, r^y_{mn}\right] \delta(\omega_{mn} - \omega). \tag{M4}$$

In the cubic chiral point group of RhSi (space group $P2_13$, point group $T$), the three-fold rotations about the cubic body diagonals enforce $\beta_{xx}(\omega) = \beta_{yy}(\omega) = \beta_{zz}(\omega)$, so the trace reduces to $\beta(\omega) = 3\,\beta_{xx}(\omega)$ and is determined by a single independent quantity. This symmetry equivalence underlies the use of $\beta(\omega)$ as the single observable in the main text.

For an isolated band crossing carrying a Berry-monopole charge $C$, the trace is fixed by the topological charge throughout the photon-energy window in which only that crossing contributes $\beta(\omega) = C\beta_0$ with $\beta_0 = \pi e^3/h^2$. In chiral multifold semimetals such as RhSi, the multifold node at Γ carries $C = \pm 2$ (depending on enantiomer), and a finite photon-energy window in which only the Γ-point transitions contribute therefore yields $\beta(\omega) \simeq 2\beta_0$. This is the quantization condition tested in Fig. 3 of the main text.

**First-principles calculation of $\boldsymbol{\beta(\omega)}$.** The CPGE tensor of $Rh_{0.95}Ni_{0.05}Si$ was evaluated from the ab initio band structure used to interpret our ARPES data. The Fermi level was fixed at the value determined experimentally by ARPES, with no adjustable parameter. Equations (M2) to (M4) were evaluated on a dense Brillouin-zone mesh using a Wannier-interpolated Hamiltonian. For the band-resolved analysis (Fig. 4b and 4d), the momentum-space integrand of $\beta(\omega)$ was retained without integration to identify the regions of the Brillouin zone that contribute to the plateau.
In Ni-doped RhSi, the low-energy CPGE response is dominated by optical transitions near the multifold fermion at Γ, producing a nearly quantized plateau over a finite photon-energy window set by Pauli blocking at the engineered Fermi level. As $\omega$ increases, transitions near the R-point multifold node, which is the opposite-chirality partner of the Γ-point multifold, become optically allowed and contribute with opposite sign, suppressing and ultimately reversing the total response, consistent with the vanishing of the net Berry-monopole charge across the Brillouin zone.

**Acknowledgements.** Los Alamos National Laboratory (LANL)-led THz emission spectroscopy and FTIR studies were supported by the LANL Laboratory Directed Research and Development (LDRD) program (Award Nos. 20240037DR, 20230014DR, 20261283IR, and 20260116ER; T.N., P.P.). Princeton-led synchrotron based ARPES measurements and supplementary ultrafast spectroscopy were supported by the U.S. Department of Energy (DOE) under the Basic Energy Sciences program (Grant No. DOE/BES DE-FG-02-05ER46200; M. Z. H.), National Quantum Information Science Research Centers, Quantum Science Center (at ORNL) supported by U.S. DOE and the Gordon, and the Gordon and Betty Moore Foundation (No. GBMF9461; M. Z. H). Synchrotron radiation experiments were performed at the BL25SU of SPring-8 with the approval of the Japan Synchrotron Radiation Research Institute (JASRI) (Proposal No. 2018A1684, No. 2019A1696, and No. 2023A1611). We acknowledge the Paul Scherrer Institute, Villigen,

Switzerland, for provision of synchrotron radiation beamtime at the ADRESS beamline of the Swiss Light Source (Proposal No. 20222089). The authors also wish to thank J. D. Delinger at Beamline 4.0.3 of the Advanced Light Source for the preparation procedure and obtaining the vacuum ultraviolet APRES results. Work was partially performed at the Center for Integrated Nanotechnologies, an Office of Science User Facility operated for the U.S. DOE Office of Science. LANL, an affirmative action equal opportunity employer, is managed by Triad National Security, LLC for the U.S. DOE National Nuclear Security Administration, under contract no. 8923318CNA000001.

**Author contributions.** B.K., T.A.C., G.C., P.P., and M.Z.H conceived the project. The THz emission spectroscopy experiments were performed by B.K., T.N., T.A.C, Z.-J.C., and P.P. FTIR measurements were carried out by B.K., T.N., A.H., and S.S., under the supervision of P.P. ARPES experiments were performed by T.A.C., Z.-J.C., X.Y., X.L., and I.B. with support from V.N.S. The crystals were grown by K.M. under the supervision of C.F. M.Y and G.C performed the first-principles calculations. Y.-X.J., P.K., M.S.H, and R.P.P. contributed to data interpretation. B.K., T.N., M.Y. G.C., M.Z.H., and P.P. wrote the manuscript with input from all the co-authors. G.C., M.Z.H., and P.P supervised the overall project.

**Competing interests**

The authors declare no competing interests.

**Data availability**

The datasets supporting the results of this study are available from the corresponding authors upon reasonable request.

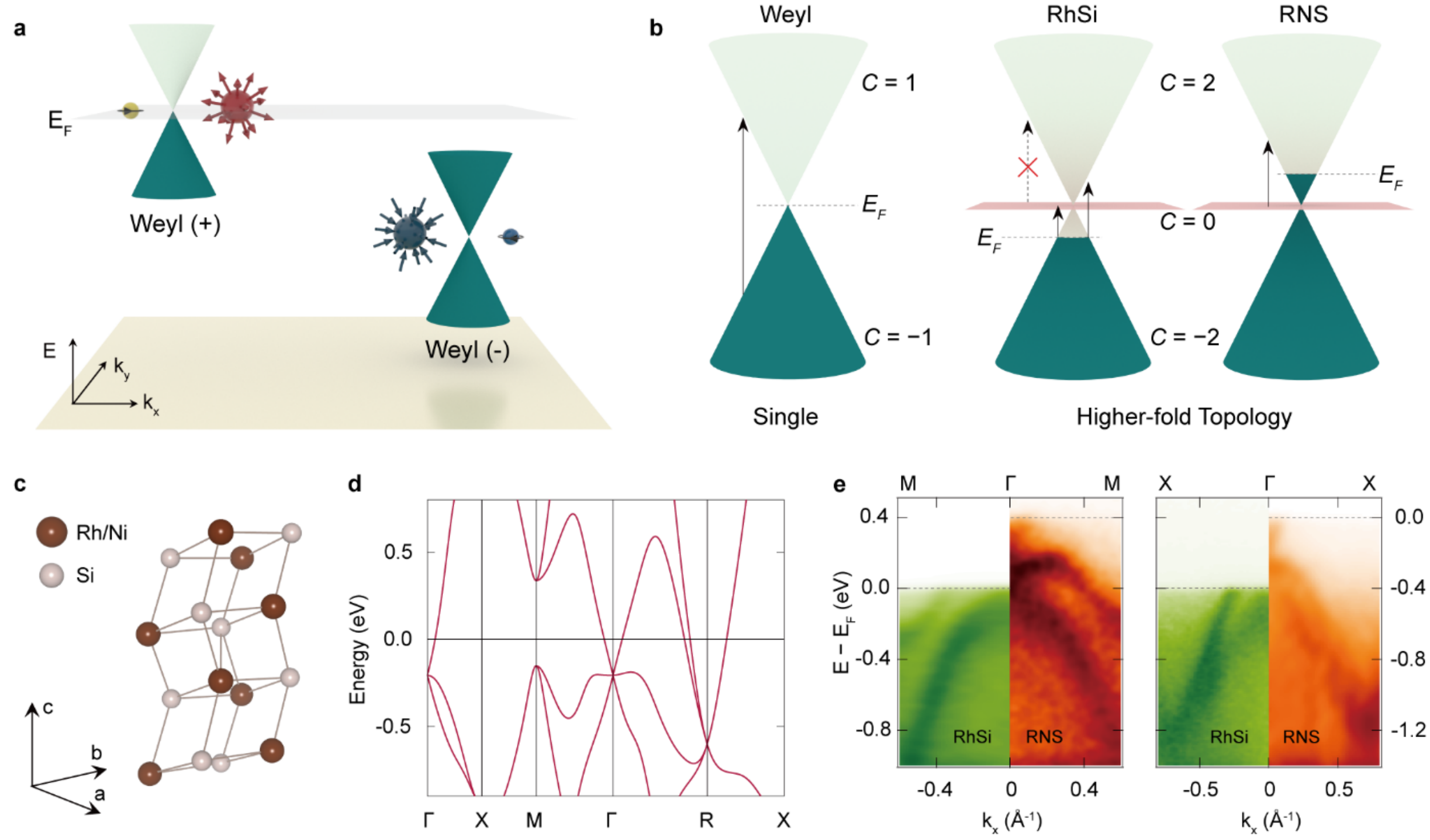

**Figure 1 | Band engineering to isolate a single Berry monopole in Ni-doped RhSi. a,** Schematic of two Weyl nodes of opposite chirality at different energies. When the Fermi level lies between the nodes, interband transitions near the lower node are Pauli blocked, opening a spectral window governed by a single Berry monopole. **b,** Comparison between an ideal single Weyl node and the multifold nodes realized in RhSi and $Rh_{0.95}Ni_{0.05}Si$ (RNS). Solid arrows and the dashed arrow denote allowed and forbidden interband transitions, respectively. **c,** Crystal structure of $Rh_{0.95}Ni_{0.05}Si$, which preserves the chiral cubic symmetry of the parent compound RhSi. **d**, DFT band structure of $Rh_{0.95}Ni_{0.05}Si$. The Γ-point multifold node lies just below the Fermi level, whereas its opposite-charge counterpart near the $R$-point is located farther from $E_F$, leading to the band alignment required for a spectrally isolated optical response. **e,** ARPES spectra of stoichiometric RhSi and $Rh_{0.95}Ni_{0.05}Si$ along with the $\mathrm{M}-\Gamma-\mathrm{M}$ (left) and $\mathrm{X}-\Gamma-\mathrm{X}$ (right) direction, confirming the predicted Fermi-level shift and the resulting single-node optical regime.

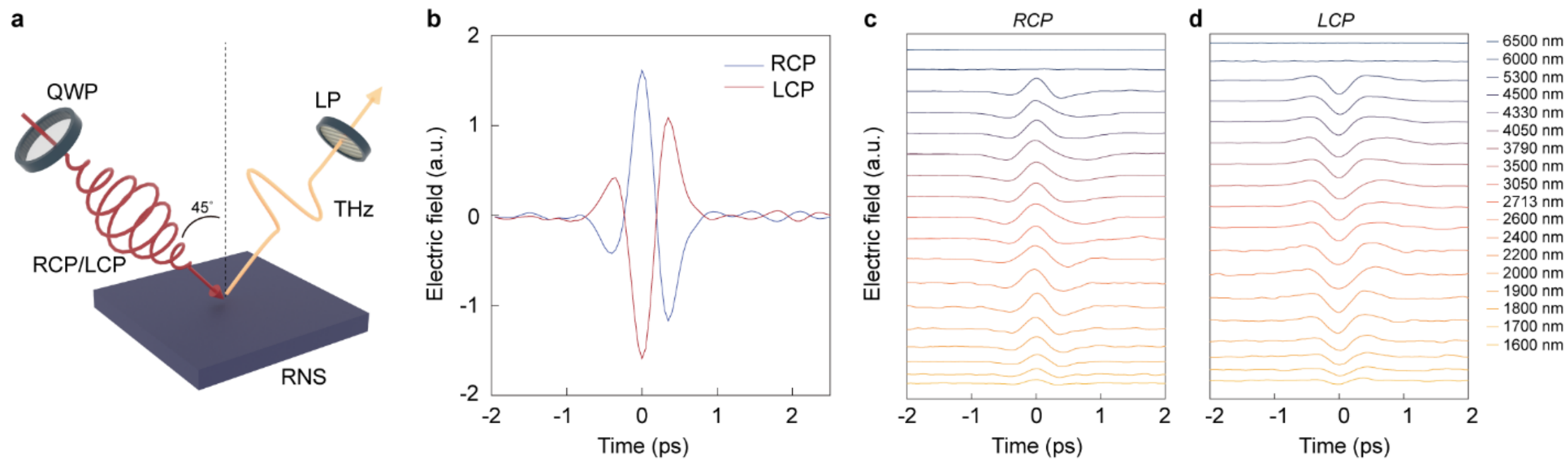

**Figure 2 | Helicity- and wavelength-dependent THz emission from Ni-doped RhSi. a,** Schematic of the THz emission geometry where circularly polarized near- and mid-infrared pump pulses are incident at 45°on a [001]-oriented $Rh_{0.95}Ni_{0.05}Si$ crystal, generating a longitudinal CPGE current with a finite in-plane projection that radiates into the far field. **b,** THz electric-field waveforms measured under 2200 nm excitation with RCP (blue) and LCP (red) pump pulses with nearly identical temporal profiles but opposite polarity, demonstrating the helicity-odd character of the emitted field. **c,d,** Wavelength-dependent THz emission waveforms measured under **c,** RCP and **d,** LCP excitation showing that the helicity-dependent emission is confined to a finite spectral window, characterized by a sharp onset, a wavelength independent intermediate regime and an abrupt long-wavelength cutoff.

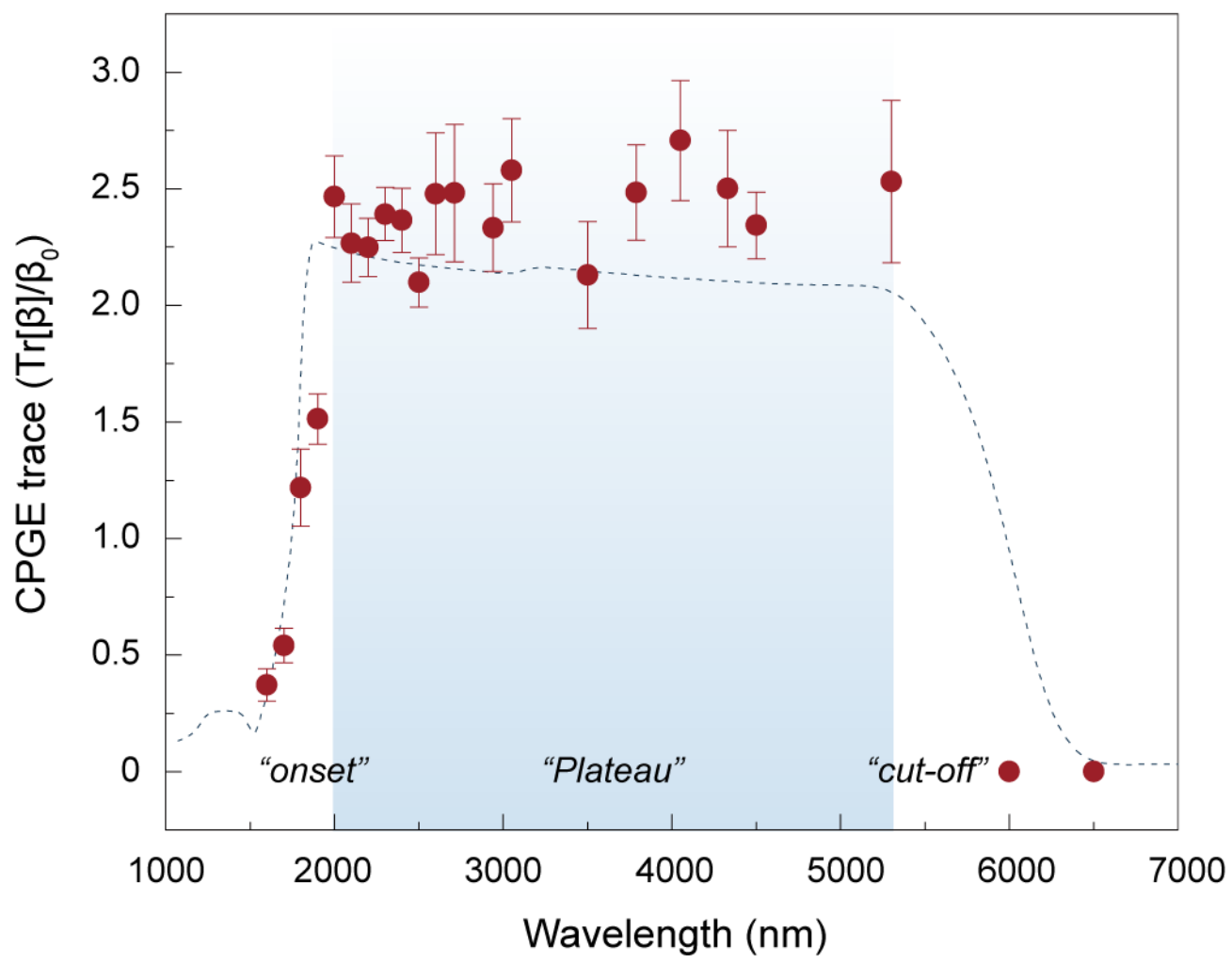

**Figure 3 | CPGE spectrum in the monopole-isolated optical window.** Wavelength dependence of the calibrated helicity-dependent CPGE trace, expressed as $\mathrm{Tr}[\beta]/\beta_0$, where $\beta_0 = \pi e^3/h^2$. Red circles denote the experimentally calibrated response extracted from the helicity-odd THz emission waveforms, and the blue dashed curve shows the first-principles CPGE calculation for the engineered $Rh_{0.95}Ni_{0.05}Si$ band alignment. The response rises sharply at the short-wavelength onset, forms a broad mid-infrared plateau near the quantized CPGE magnitude, and collapses at the long-wavelength cutoff. The shaded region marks the plateau window in which interband transitions are dominated by the chemically shifted Γ-point multifold node with $C_{\mathrm{exp}} \simeq 2.39$. The agreement between the calibrated plateau and the calculated response supports a Berry-monopole-governed CPGE injection current rather than an unconstrained nonlinear optical background.

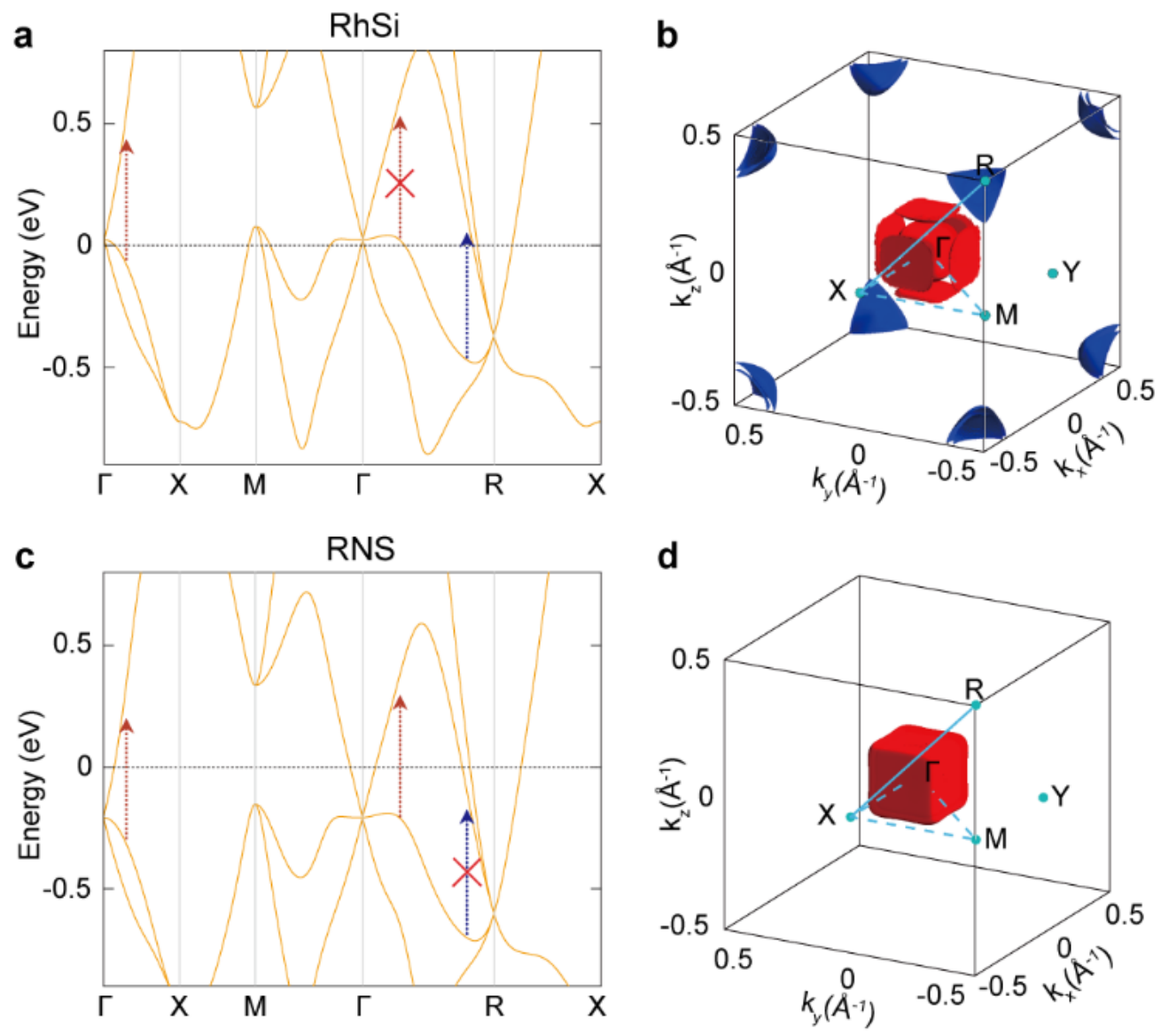

**Figure 4 | First-principles assignment of the monopole-isolated optical window. a,** Calculated bulk band structure of stoichiometric RhSi along the high-symmetry directions. The Fermi level, shown by the dashed line, intersects the low-energy branches of the zone-center multifold node, while interband transitions associated with the opposite-charge node near the $R$-point remain accessible within overlapping photon-energy ranges. Red and blue arrows denote representative transitions associated with the Γ- and R-point multifold nodes, respectively; the red cross marks a Pauli-blocked transition channel at the stoichiometric Fermi level. **b,** Momentum-resolved CPGE contribution of stoichiometric RhSi at $\hbar\omega = 450$ meV. Red and blue regions denote contributions from transitions near Γ and R, respectively, showing that opposite-charge channels coexist within the same optical window and prevent a clean single-node response. **c,** Calculated bulk band structure of $Rh_{0.95}Ni_{0.05}Si$. Electron doping shifts the band manifold such that the Γ-point multifold node lies below $E_F$, while the competing R-point transitions are Pauli-blocked and excluded from the plateau window, as indicated by the red cross. **d,** Momentum-space distribution of CPGE contributions, revealing a compact region centered at the Γ-point that governs the response and underlies the plateau measured in Fig. 3.

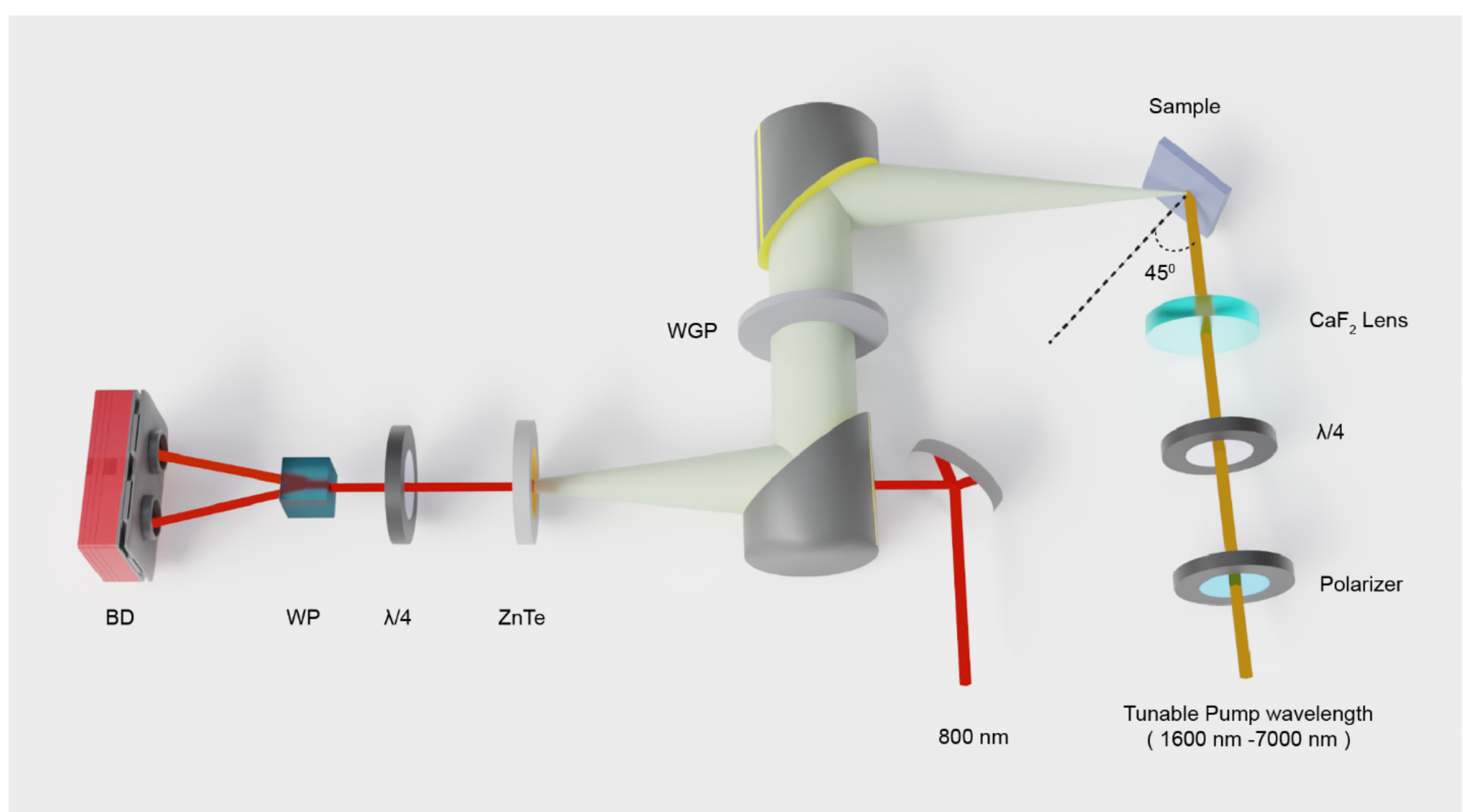

**Extended Data Fig. 1 | Schematic of the THz emission experimental setup.** The sample is excited by a wavelength-tunable pump beam incident at 45° with respect to the surface normal. The pump helicity is controlled using a polarizer and a QWP. The emitted THz transient is collected by parabolic mirrors and focused onto a ZnTe crystal for electro-optic sampling, where it is temporally resolved using a time delayed optical gate beam. BD: balanced photodetector, WP: Wollaston prism; WGP: wire-grid polarizer

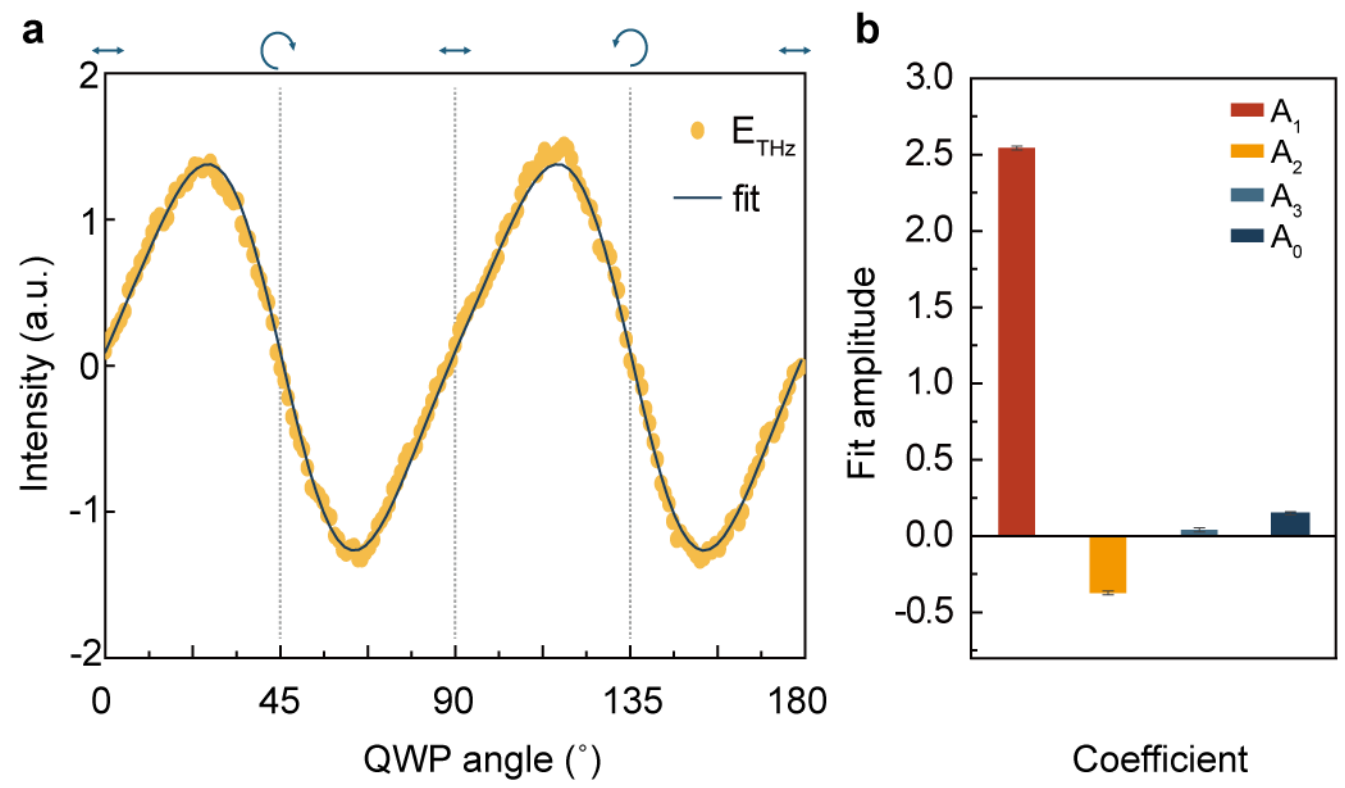

**Extended Data Fig. 2 | Polarization symmetry of the THz emission plateau. a**, QWP-angle dependence of the peak THz field under 2200 nm excitation, recorded at a fixed time delay corresponding to the waveform maximum. Yellow symbols are the measured field amplitudes, and the solid line is a fit to Eq (1). **b**, Fitted amplitudes of the symmetry components in panel **a**. The response is dominated by the $A_1$ coefficient, while the linear-polarization-dependent ($A_2$ and $A_3$) and polarization-independent ($A_0$) terms are much smaller, supporting the CPGE origin of the THz emission plateau.

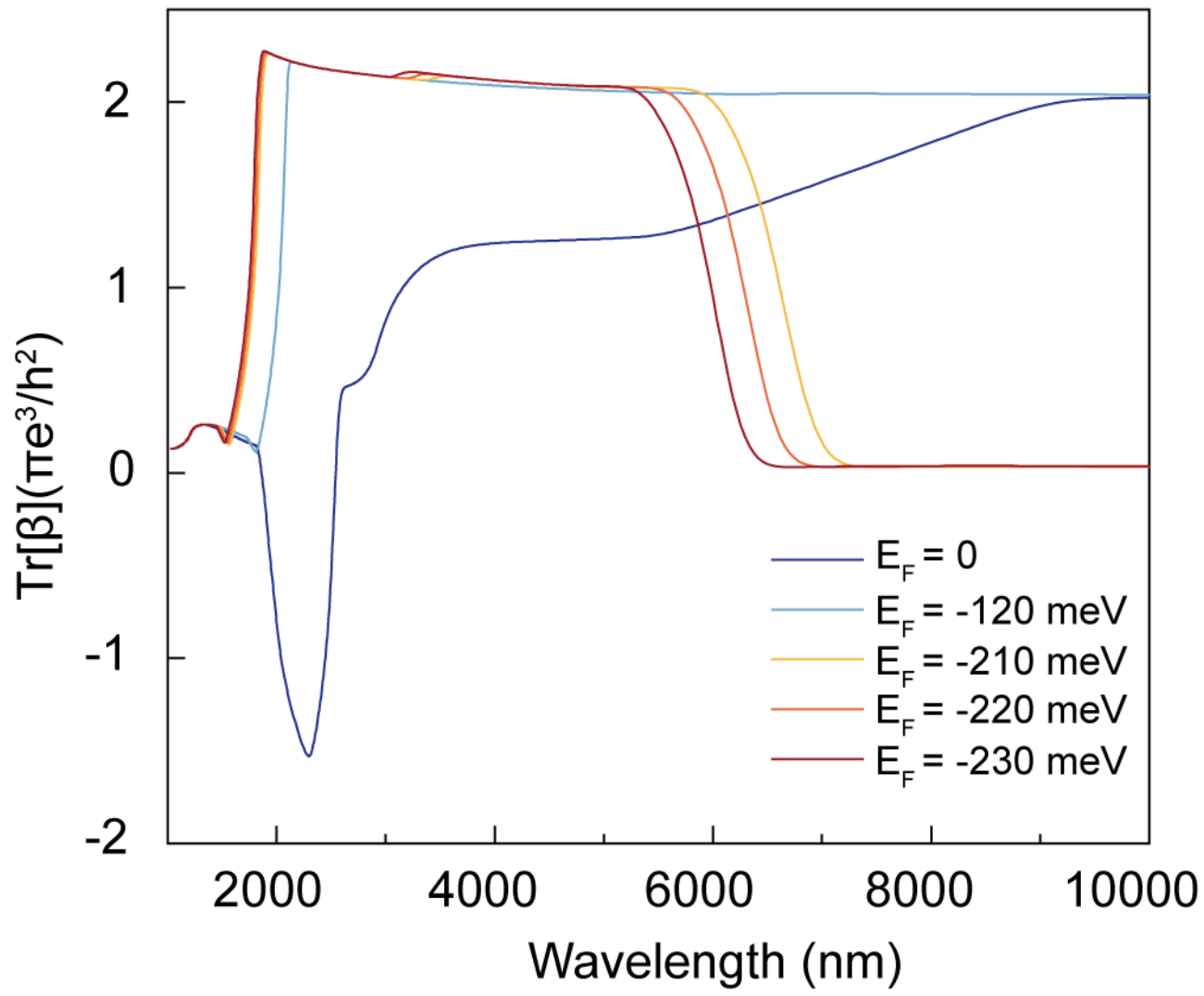

**Extended Data Fig. 3 | Chemical-potential evolution of the calculated CPGE response.** Calculated trace of the CPGE tensor, $\mathrm{Tr}[\beta(\omega)]$, as a function of pump wavelength for a series of rigid downward shifts of the pristine RhSi band manifold. The label $E_F = 0$ denotes the response obtained from the intrinsic RhSi band alignment, and each negative value gives the rigid energy by which the band manifold is displaced below the intrinsic calculation, approximating the chemical-potential shift produced by Ni substitution. At the intrinsic alignment ($E_F = 0$), $\mathrm{Tr}[\beta]$ is strongly wavelength dependent and reverses sign near 2400 nm, reflecting the competing opposite-charge contributions of the Γ- and R-point multifold nodes. As the manifold is shifted downward, the sign reversal is suppressed and a broad positive plateau emerges near the $C \simeq 2$ value; by a shift of about 220 meV, consistent with the alignment in our crystal established by ARPES, the plateau spans the full mid-infrared window before terminating at a sharp long-wavelength cutoff. The cutoff moves to shorter wavelength as the shift increases, tracking the Pauli-blocking edge of the engineered band alignment. The near-quantized plateau therefore appears only when chemical tuning places the system within a single-monopole-isolated optical window.